\documentclass[a4paper,fleqn,usenatbib]{mnras}

\usepackage[T1]{fontenc}
\usepackage{ae,aecompl}

\usepackage{graphicx}
\usepackage{subcaption}
\usepackage{url}
\usepackage{amsmath} 
\usepackage{float}

\title[Gamma-ray Novae: Rare or Nearby?]{Gamma-ray Novae: Rare or Nearby?}

\author[P. J. Morris et al.]{
Paul J. Morris,$^{1}$\thanks{E-mail: paul.morris@physics.ox.ac.uk (PJM)}
Garret Cotter,$^{1}$
Anthony M. Brown$^{2}$
and Paula M. Chadwick$^{2}$
\\
$^{1}$Oxford Astrophysics. Denys Wilkinson Building, Keble Road, Oxford, OX1 3RH, United Kingdom\\
$^{2}$Department of Physics and Centre for Advanced Instrumentation, Durham University, Durham, DH1 3LE, United Kingdom\\
}

\date{Accepted XXX. Received YYY; in original form ZZZ}

\pubyear{2016}

\begin{document}
\label{firstpage}
\pagerange{\pageref{firstpage}--\pageref{lastpage}}
\maketitle

\begin{abstract}
Classical Novae were revealed as a surprise source of $\gamma$-rays in {\sl {\sl Fermi}} LAT observations. During the first 8 years since the LAT was launched, 6 novae in total have been detected to $ > 5\sigma$  in $\gamma$-rays, in contrast to the $69$ discovered optically in the same period. We attempt to resolve this discrepancy by assuming all novae are $\gamma$-ray emitters, and assigning peak one-day fluxes based on a flat distribution of the known emitters to a simulated population. To determine optical parameters, the spatial distribution and magnitudes of bulge and disc novae in M31 are scaled to the Milky Way, which we approximate as a disc with a $20~\rm{kpc}$ radius and elliptical bulge with semi major axis $3~\rm{kpc}$ and axis ratios 2:1 in the xy plane. We approximate Galactic reddening using a double exponential disc with vertical and radial scale heights of $r_{d} = 5~\rm{kpc}$ and $z_{d} = 0.2~\rm{kpc}$, and demonstrate that even such a rudimentary model can easily reproduce the observed fraction of $\gamma$-ray novae, implying that these apparently rare sources are in fact nearby and not intrinsically rare. We conclude that classical novae with $m_{R} \leq 12$ and within $\approx 8~\rm{kpc}$ are likely to be discovered in $\gamma$-rays using the {\sl {\sl Fermi}} LAT.
\end{abstract}

\begin{keywords}
Novae -- gamma-rays -- cataclysmic variables -- extinction -- M31
\end{keywords}



\section{Introduction}

Cataclysmic variables (CVs) are semi-detached binary systems consisting of a white dwarf accreting from a lower mass stellar companion which has overfilled its Roche Lobe. They are progenitors for nova events, the most luminous and therefore most easily-detectable sub-class of which are the classical novae (CNe, CN singular). Such events are characterised by a typical increase in optical luminosity of a factor of $10^{6}$ \citep{CO:2007}, powered by a thermonuclear runaway (TNR) on the surface of the white dwarf (e.g. \citet{Shara:1989}). Though $\gamma$-rays were hypothesised to arise from the beta decay of proton-rich elements produced in the TNR by \citet{CH:1974}, these were predicted to be in the $\sim$ 1\,MeV range, hence it came as something of a surprise when on 10 March 2010 the CV V407 Cyg was detected in $\gamma$-rays using the {\sl Fermi} Large Area Telescope (LAT) during a CN outburst \citep{2010Sci...329..817A}. Due to the unusual nature of the Mira variable containing V407 Cyg, \citet{2010Sci...329..817A} hypothesised that the $\gamma$-ray emission arose as a consequence of the strong stellar wind absent from more typical CN systems, and concluded that $\gamma$-ray CNe would be exceptionally rare. Just over two years later V1324 Sco became the second CN observed in $\gamma$-rays \citep{SCOdisc:2012}, which has now been joined by V959 Mon \citep{2012ATel.4310....1C}, V339 Del \citep{Cheung:2013b}, V1369 Cen \citep{2013ATel.5653....1C} and V5668 Sgr \citep{novaSgr2015} in being observed to more than $5\sigma$ certainty (See \citet{4novae:2014} and \citet{2novae:2016} for a complete summary). The $\gamma$-ray novae all exhibit very similar light curves. 

In contrast, in the first 8 years since the LAT first began taking data in August 2008, a total of 69 \citep{MWlist} \footnote{Accessible at http://asd.gsfc.nasa.gov/Koji.Mukai/novae/novae.html} novae have been discovered optically. Many reasons have been put forward to explain this discrepancy, with one possibility being that we are only able to detect $\gamma$-rays from novae occurring close to the solar neighbourhood. Although few CNe have robust distance measurements, distance estimates to all the identified detected $\gamma$-ray novae place them within 4.5 kpc, which supports the notion that they are all relatively nearby within the Milky Way  \citep{4novae:2014,2novae:2016}. The same authors note that with the exception of V407 Cyg, there is nothing to indicate any of the $\gamma$-ray novae are particularly unusual. Another possibility is that we can only observe the most luminous $\gamma$-ray novae. Additionally, as is likely the case in V407 Cyg \citep{2010Sci...329..817A}, such phenomena may be driven by unusual conditions in the local environment which can accelerate particles to the high energies required to produce $>100$ MeV photons.

In this paper, we investigate the apparent rarity of $\gamma$-ray novae by simulating a Galactic nova population using novae in M31 to determine their optical properties and the Galactic $\gamma$-ray novae for their corresponding high-energy ones. 

\begin{table*}
\centering
\caption{Table listing key properties of the $\gamma$-ray detected novae based on the daily bin with the maximum $TS$ value. See \citet{4novae:2014}, \citet{2novae:2016} and the contained references for more information on daily binned light curves and the V1369 Cen distance. The adopted distances to V407 Cyg and V1324 Sco are inferred from estimating the line of sight extinction relative to a RC star and V959 Mon from expansion parallax \citep{novae_distances}. \citet{Chocol:2014} inferred the V339 Del distance from the maximum-magnitude rate of decline relation (eg \citet{1985ApJ...292...90C})  and \citet{2016MNRAS.455L.109B} use infra-red emission from the nova shell of V5668 Sgr to infer its distance. V1369 Cen currently has no more reliable distance estimate. $F_{GalDiff}$ is the flux attributed to the Galactic diffuse on the sky pixel spatially coincident with the position of each nova. The $TS$ values correspond to the peak daily flux.}
\begin{tabular}{c | c c c c c c}
nova & V407 Cyg & V1324 Sco & V959 Mon & V339 Del & V1369 Cen & V5668 Sgr \\
\hline
Peak daily flux, $F_{\gamma}$ (10$^{-7}$ ph s$^{-1}$ m$^{-2}$) & 13.9$\pm$2.6 & 12.3$\pm$2.9 & 13.8$\pm$3.7 & 5.9$\pm$1.1 & 5.1$\pm$1.3 & 1.8$\pm$0.8 \\
$F_{\gamma}/F_{GalDiff}$ & 0.254 & 0.185 & 0.305 & 0.381 & 0.0897 & 0.0704  \\
$TS$ value & 56.8 & 35.0 & 27.7 & 65.7 & 37.6 & 11.6 \\
Distance (kpc) & 3.5$\pm$0.3 & 4.3$\pm$0.9 & 2.3$\pm$0.6 & 3.2$\pm$0.3 & 2.5 & 1.5$\pm$0.2 \\
\hline
\end{tabular}
\label{nova_table}
\end{table*}

\section{The $\gamma$-ray Novae}\label{GRN}

\subsection{V407 Cyg}
First observed in 1936 during an outburst \citep{V407:1949}, V407 Cyg is a relatively well-studied system. It belongs to a rare subgroup of CVs known as symbiotic Miras, in which the secondary is pulsating red giant (RG) known as a Mira variable. The WD accretes from the RG wind rather than via an accretion disk. This distinction was made for the first time in 1994 by \citet{Mira:1998}, whilst \citet{1966IBVS..125....1M} deduced the period of the Mira pulsations to be $745^{d}$, the phase of which determines the optical magnitude of the system, which typically resides between $m_{V}=13.5-17$. Miras are believed to be surrounded by a dust envelope, a fact used by \citet{Munari:1989} to attribute a sine wave superimposed on the B-band peaks to the orbital period of the system. They speculate that ionizing radiation from the WD inhibits the dust formation except in a shadow cone-shaped region produced by the RG which causes a wavelength dependent shift to be observed at various orbital phases. Munari et al. concluded that the orbital period was $P=43\pm5yr$. A second nova-like flare occurred in 1998 \citep{Kolotilov:2003} lasting until 2002, with a peak magnitude of around 11 attained. It is unknown whether any $\gamma$-rays would have been produced. 

V407 Cyg underwent its most recent nova outburst on 10 March 2010, the  optical magnitude of which reached $m_{V}\approx8$ at its peak. It was during this event that $\gamma$-rays were observed \citep{2010Sci...329..817A}, implying the two were related. \citet{2010Sci...329..817A} found that a $\gamma$-ray transient detection using the LAT was consistent with the established optical location of V407 Cyg with only a 0.040$^{\circ}$ offset, giving a 95\% chance that the $\gamma$-rays had indeed originated from it, and crucially that no other high energy sources were in the error circle.  As V407 Cyg is an exceptional symbiotic system for which the proposed $\gamma$-ray emission mechanisms appeared related to its unusual nature, \citet{2010Sci...329..817A} concluded that the emission of $\gamma$-rays from CVs would be extremely rare. 

\subsection{V1324 Sco}

V1324 Sco was discovered on May 22 2012 with an I band magnitude of $m_{I}=19.5$ \citep{Wagner:2012}, brightening to $m_{I}=11$ by 2 June. The optical peak of $m_{V}\approx10$ occurred on 20 June, and slowly declined, with the time to decline by two visual magnitudes $t_{2}\approx25$ days \citep{Cheung:econf}. From spectroscopic evidence, V1324 Sco is reminiscent of a Fe-II classical nova (compared to V407 Cyg, a He/N nova \citep{Cheung:econf}). Such systems have strong Fe-II lines present in their spectra thought to originate from interactions of the nova shell with a gas envelope from the secondary companion. Their presence is tied to the evolution of the secondary star \citep{Williams:2012}. 

The discovery of a {\sl Fermi} LAT detected transient at a location consistent with the optical location of V1324 Sco confirmed it as the first CN source of $> 100$~MeV $\gamma$-rays. This came as something of a surprise as the possible $\gamma$-ray emission mechanisms all appeared to be linked to the dense RG wind, not thought to be present in classical novae.

\subsection{V959 Mon}
V959 Mon was first identified as a $\gamma$-ray transient on 22 June 2012, making it the first nova for which the $\gamma$-ray discovery preceded the optical \citep{2012ATel.4310....1C}. This was largely as a result of its apparent close proximity to the Sun ($\approx 20^{\circ}$) during the classical nova outburst, consequently the peak optical magnitude and $t_{2}$ are unknown, and optical confirmation of the nova was only obtained in August 2012 \citep{2012CBET.3202....1F} when $m_{V}=9.4$. 

\citet{Shore:2013} concluded V959 Mon was an oxygen-neon nova by looking at the available spectroscopic data from around 55 days after the outburst. The overabundance of oxygen, neon and magnesium present in their ejecta are thought to originate from pre-outburst enrichment of the envelope of a white dwarf of mass close to the Chandrasekhar limit. Periodic oscillations observed in multiple wavelengths have been confirmed as orbital \citep{Osborne:2013}, making V959 Mon the only $\gamma$-ray nova with orbital inclination along the line of sight ($P_{ORB}=0.2957\pm{0.0007}^{d} \approx 7.10\pm{0.02}$~hr).

\begin{figure}
\includegraphics[width = 8.0cm, keepaspectratio=true]{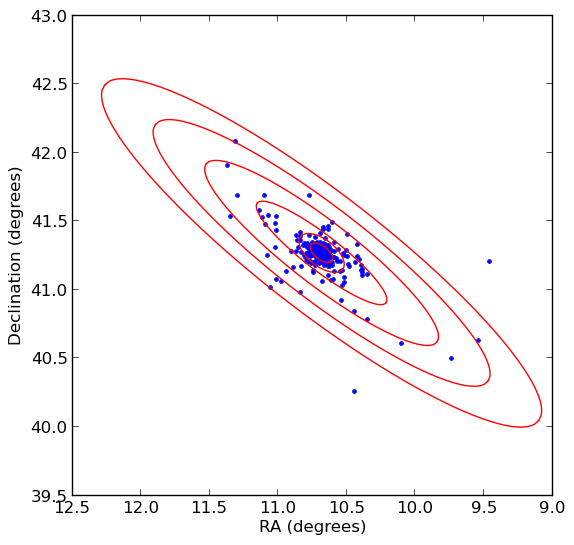}
\caption{Figure demonstrating the spatial elliptical bins used for the novae in M31. Novae are shown as blue points, with the red ellipses showing the spatial bin boundaries. The inner 2 bins define the bulge, and have a different $a/b$ compared to the outer 4 disc bins. Any novae outside of the largest ellipse were considered external to M31 and subsequently excluded. The inclination angle of the M31 semi-major axis relative to a line of constant declination is taken to be $37.7^{\circ}$ \protect\citep{1958M31size}.}
\label{EllipticalBins}
\end{figure}

\subsection{V339 Del}
Discovered on 14 August 2013 by \citet{2009CBET.2050....1Y} at an optical magnitude of $6.8$, V339 Del reached a maximum brightness of $m_{V}=4.43$  around 2 days later \citep{2013IBVS.6080....1M} and was visible to the naked eye. A measurement at $m_{V} \approx 17.1$ on 13 August 2013 implies a very fast rise to maximum, and \citet{Chocol:2014} measured a fast decline, with the time to decline by two visual magnitudes $t_{2}\approx 10$~days.  

\subsection{V1369 Cen}

V1369 Cen was discovered by \citet{Cen_discovery:2013} on 2 December 2013 but reached a first optical maximum of $m_{V} = 3.6$ (AAVSO\footnote{American Association of Variable Star Observers (\url{http://aavso.org/lcg})}) three days later. The onset of $\gamma$-ray emission coincided with a second optical maximum \citet{2013ATel.5653....1C}, and \citet{2013ATel.5639....1I} inferred that the nova is relatively nearby by considering equivalent widths of of the Na I doublet to estimate the extinction. \citet{2014ATel.6413....1S} also use spectra to estimate a distance of $\approx 2.4~\rm{kpc}$.

\subsection{V5668 Sgr}

The most recent confirmed $\gamma$-ray novae was discovered on 15 March 2015 by \citet{2015CBET.4080....1S}. Much like V1369 Cen, the visual AAVSO light curve exhibits multiple optical peaks with the first maximum at $m_{V}=4.1$ \citep{2novae:2016}. \citet{2016MNRAS.455L.109B} infer a distance of $d = 1.54~\rm{kpc}$ from measuring the expansion parallax of the nova shell, and argue that the multiple optical peaks are a manifestation of strong dust production caused optical emission to be re-radiated in the infra-red and the exact geometry of the nova shell allowing some optical light to escape.

\section{Galactic nova Rate}

In order to simulate novae in the Milky Way, it is necessary to estimate their occurrence rate. The location of the Solar System within the Galactic disc complicates matters, as optical emission is scattered from dust grains in the interstellar medium (ISM). This causes interstellar extinction, making it impossible to view every nova in the Milky Way typically reducing us to $\approx 10$ nova detections annually. In the past, estimates have ranged enormously from $11$ to $260$~yr$^{-1}$ (\citet{Shafter:1997} and the references therein), demonstrating that deducing such a rate is non-trivial. 

When attempting to deduce a nova rate, one of two approaches is typically taken. The first concerns Galactic data, in which distances are deduced to nearby novae and combined with assumptions regarding their spatial distribution, which manifest themselves as high uncertainties. Unless a CN occurs close enough that the nova shell can be spatially resolved allowing a distance to be inferred (e.g. \citet{Monshell:2013}), novae distances are notoriously difficult to measure and often involve the assumption that novae are standardisable candles \citep{1985ApJ...292...90C}. Deductions typically agree with the values of $29\pm 17$~yr$^{-1}$ derived by \citet{Ciardullo:1990} or the $35\pm 11$~yr$^{-1}$ deduced by \citet{Shafter:1997}. Conversely, \citet{Liller:1987} estimated a rate of $73\pm 24$~yr$^{-1}$, demonstrating the uncertainties present in this method.

An alternative approach is to consider extragalactic nova populations, and scale them to the Milky Way by using, for example, the mass to light ratio. An example of this is \citet{dellaValle}, who infer a nova rate of 20 yr$^{-1}$, consistent with the lower end of the Galactic procedure. An advantage of this method is that a much larger sample of the nova population can be observed in a nearby galaxy, such as M31. Novae here are also approximately equidistant and can be assumed to have similar reddening along the line of sight.

For this work, we leave our Milky Way nova rate as a free parameter consistent with $\dot{N}_{novae} = 35 \pm 11 \ \rm{year}^{-1}$ to test whether a compatible rate is capable of reproducing the observed nova rate.

\section{CNe Population in M31}

Due to the advantages outlined above, it was decided to use information from an extragalactic nova population to determine the spatial distribution and optical luminosities of our simulated novae, with M31 being the obvious candidate on which to model our nova population due to its close proximity. A list of all observed novae in M31 dating back to 1909 is available (\citet{M31list}\footnote{Accessible at: {http://www.mpe.mpg.de/~m31novae/opt/m31/index.php}}, \citet{M31cat:2007} and \citet{M31cat:2010} and the references therein). To account for the relative orientation of Andromeda with respect to the line of sight, spatial binning was done elliptically, and defined differently for the disc and bulge regions. 

The bulge-disc boundary and ratio of semi-major to semi-minor axes, $a/b$, were defined according to isophotes detailed in \citet{Beaton:2007}. Hence we adopt $a_{boundary}=700''$, corresponding to a physical distance of 3 kpc considering the M31 distance of 780 kpc \citep{M31dist}. Due to large number of bulge novae in M31, the bulge region was further subdivided into two sections each with the same $a/b$, with the inner-outer bulge boundary corresponding to $a=350''$. The number of novae in each bin are subject to uncertainties caused by projection effects. These effects are not constant for each bin, but increase with the size of the semi major axis, $a_i$, and scale height of the bin. Novae within the M31 disc are likely to be close to the galactic plane, hence the dominant disc uncertainty is $a_i$. Bulge novae are likely to exhibit a larger range of heights above the plane; however the larger sample and apparent symmetry of M31 should mitigate this effect. 

For the disc, the ratio $a_{disc}/b_{disc}$ was defined based on the inclination angle of $i=12.5^{\circ}$ \citet{1978M31inc}, with the maximum semi major axis of $a=4.0^{\circ}$ consistent with $2.0^{\circ}$ the observed angular extent of M31 \citep{1958M31size}. The disc region was subsequently divided into four linearly spaced sections, with the semi-minor axes determined as $b=a \cos i$. Four sections were chosen as they allowed a better spatial distribution to be determined whilst containing multiple novae per bin. Binning is shown in Fig. \ref{EllipticalBins}.

Data from multiple filters were available, but due to being recent (oldest data from 1990), having a large sample size (274) and being less affected by interstellar extinction, it was decided to focus on R-band data. This list also includes recurrent novae (RNe), which are defined as those which have been observed multiple times with periods of quiescence that can last years to decades. RNe which have only been observed in outburst once are difficult to distinguish from CNe, although RNe are typically 1000 times less luminous \citep{CO:2007}. As $\gamma$-rays have not been detected from any typical RNe, they should be emitted from any sample, and this was achieved by removing any novae with coordinates spatially consistent with other nova events. In total, 12 RNe were omitted, leaving 262 (176+86 = bulge + disc) novae. This is consistent with the result obtained by \citet{2015ApJS..216...34S} who estimate $\approx 4\%$ of nova events in M31 are RNe. Dereddening corrections were applied by using the NASA/IPAC Extragalactic Database\footnote{https://ned.ipac.caltech.edu} value of $A_B=0.300$ magnitudes which is based on HI column densities \citep{HI:1982}, with $A_R=(2.32/4.10) A_B $ using the mean extinction curves in \citep{SM:1979}. The absolute magnitude, $M_R$, of each nova could then be found, taking the  distance to M31 to be 780 kpc.

\begin{figure}
\includegraphics[width = 8.0cm, keepaspectratio=true]{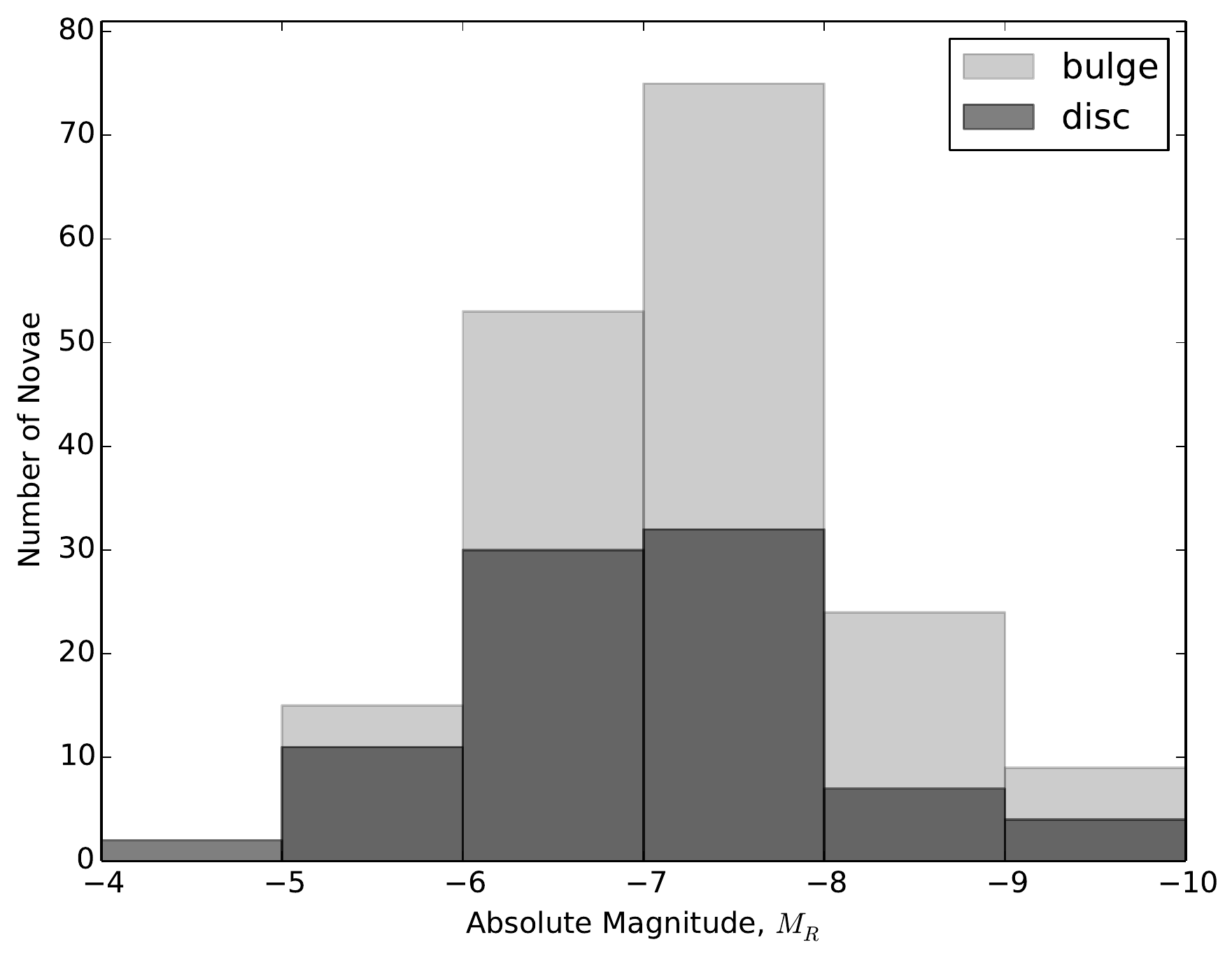}
\caption{Histogram displaying the $M_R$ values for the bulge and disc nova populations of M31. It can be seen that the two distributions are different, with a 2 sample Kolmogorov-Smirnov test indicating a 26.5\% probability that the bulge and disc populations are intrinsically the same. Recurrent novae have been omitted from the original data of \protect\citet{M31list}. With the exception of the $-5.0 < M_{R} \leq -4.0$ disc bin, each bin contains $\geq 4$ novae. }
\label{MagBins}
\end{figure}

Magnitude bin widths were chosen such that no bin was completely depleted of novae. It was decided that each bin should contain $\geq 4$ novae, with one-magnitude bin widths allowing for 5 bins under this criteria. The results are shown in Fig. \ref{MagBins}. The likelihood of M31 bulge and disc novae being sub samples of the same population was assessed using a 2 sample Kolmogorov-Smirnov (KS) test, and we cannot reject the null hypothesis that the populations are same to lower than 26.5\%. Even so, the assignment of absolute magnitudes to simulated novae was done separately for disc and bulge novae, and were based on the distribution of $M_R$ values in Fig. \ref{MagBins}. The counts per elliptical bin in Fig. \ref{EllipticalBins} and histogram in Fig. \ref{MagBins} were converted to probability distributions, such that they could be used to assign radial and $M_R$ values to a simulated nova population.

\section{Producing CNe in the Milky Way}

\subsection{Milky Way Distribution}

For simplicity, we assume novae are found either in the bulge or disc, and neglect additional Galactic components. We define a Milky Way bulge semi-major axis of $a_{b}=3.0 \ \rm{kpc}$ and disc radius of $R_{d}=20 \ \rm{kpc}$, and approximate the bulge in the xy plane as an ellipse with axis ratios 2:1 and set the angle $\phi=20^{\circ}$ between the bulge semi major axis and the vector between the Galactic centre and Solar System \citep{1997MNRAS.288..365B}. The M31 binned data were used to populate the Galactic plane with novae, with two concentric ellipses ($a:b = 1:0.5$) describing the bulge region and 4 circles to mirror the number of M31 bins. The semi major axes of these bins were obtained by normalising the areas relative to M31, such that the equivalent Milky Way disc bin contains the same fraction of the M31 disc. Finally, to account for the larger size of M31, the number of novae in each M31 bin were divided by the apparent bin area and scaled accordingly to the Milky Way. The resulting novae counts were normalised, allowing each Milky way bin to be populated with x and y positions assigned randomly but uniformly within the given bin ($a_b$ was aligned along the x axis). For disc novae, the z position was assumed to take the form $P(z) \propto {\rm exp}(-z/z_{d}) $, where $z_{d}$ is the disc scale height. We adopt $z_{d}=350 \ \rm{pc} $ \citep{DJ:1994} to mirror the old disc population from which novae derive.

To deduce $z$ positions for bulge novae, we tested several models from the literature. One was the bulge model of \citet{1997MNRAS.288..365B} for $L$-band surface brightness which we assume scales with stellar density, $\rho_B$, such that,

\begin{subequations}
\begin{equation}
\rho_{B} = \rho_0 \frac{ \rm{e}^{-a^2/a_{m}^2} }{ \left( 1+a/a_0 \right)^{1.8} },
\label{Binney1}
\end{equation}

\begin{equation}
a = \left( x^2 + \frac{y^2}{y_{0}^2} + \frac{z^2}{z_{0}^2} \right) ^{1/2},
\label{Binney2}
\end{equation}
\end{subequations}
where $\rho_0$, $a_m$, $a_0$, $y_0$ and $z_0$ were all left as free parameters. \citet{1995ApJ...445..716D} test different models for fitting the infrared surface brightness of the Galactic bulge, assessing each one with a $\chi_{\nu}^2$ fit. We test the application of some of their models to Galactic novae, namely,

\begin{subequations}

\begin{equation}
\rho_{1} = \rho_0 \exp(-0.5 r^2),
\label{Gauss}
\end{equation}

\begin{equation}
\rho_{2} = \rho_0 r^{-1.8} \exp(-r^3),
\label{powexp}
\end{equation}

\begin{equation}
\rho_3 = \rho_0 \exp(-r),
\label{exp}
\end{equation}
\end{subequations}
where $r$ is defined by,

\begin{equation}
r = \left[ \left(\frac{x}{x_0} \right)^2 + \left(\frac{y}{y_0} \right)^2 + \left(\frac{z}{z_0} \right)^2 \right]^{\frac{1}{2}},
\label{rnorm}
\end{equation}
where the parameters $\rho_0$, $x_0$, $y_0$ and $z_0$ we left free. In order to evaluate these functions, a population of novae was simulated taking $z$ values for the disc population as described previously, and using each function above to describe the bulge whilst varying the free parameters. This was done due to difficulties distinguishing between observed disc and bulge novae. Each fit was then compared to the observed sample via a 2-sample KS test, with results displayed in Table \ref{bulgefit_table}.

\begin{table} 
\centering
\caption{Table showing best fit parameters for the tested models with respect to reproducing the observed novae population. $\mathrm{p}_{KS}$ gives the probability that he simulated novae derive from the same global population as the observed novae. It can be seen that the best fit was obtained for the Gaussian model. Simulated populations were corrected for reddening effects using Eqn. \ref{redsum}.}
\begin{tabular}{c | c c c c c c c}
Model & $\rho_0$ & $x_0$ & $y_0$ & $z_0$ & $a_0$ & $a_m$ & $\mathrm{p}_{KS}$ \\
\hline
Eqn. \ref{Binney1} & 890 & - & 0.674 & 1.00 & 0.01 & 1.0 & 0.771 \\
Eqn. \ref{Gauss} & $1 \times 10^6$ & 4.17 & 0.674 & 0.344 & - & - & 0.949 \\ 
Eqn. \ref{powexp} & $1 \times 10^6$ & 0.817 & 0.838 & 0.45 & - & - & 0.893 \\
Eqn. \ref{exp} & $1 \times 10^7$ & 1.11 & 0.744 & 1.00 & - & - & 0.575 \\
\hline
\end{tabular}
\label{bulgefit_table}
\end{table}

\subsection{Milky Way reddening}

Effects due to interstellar absorption must be accounted for when considering the number of novae in our simulations that it would be possible to detect in the R-band. To do this, we apply the R-band corrected double exponential dust distribution model of \citet{DJ:1994}, such that the R-band extinction, $\alpha(r,z)$, at any point within the Milky Way in units of $\Delta m_R$~kpc$^{-1}$ along the line of site is given by, 

\begin{equation}
\alpha(r,z) = \frac {A_R}{A_V}\alpha_{GC} {\rm exp}\left( \frac{-r}{r_d} \right) {\rm exp}\left( \frac{-|z|}{z_d} \right),
\label{DJred}
\end{equation}
where $\alpha_{GC} = 9.4~m_V ~\rm{pc}^{-1}$ and $A_R/A_V = 2.32/3.1$ \citep{1999PASP..111...63F}. We assume that the spatial distribution of dust has scale height $z_d=0.2~\rm{kpc}$, and again use the argument of \citet{DJ:1994} (hereafter DJ) that the disc surface density decreases with scale distance $r_d=5~\rm{kpc}$, and assume that the Galactic dust traces this. We use the method of the same authors to compute the reddening along the line of sight to each nova in increments $\Delta s$ of no greater than 50pc, such that the total magnitude gain due to reddening effects is given by,

\begin{equation}
\Delta m_R = \sum_{i} \alpha_i \Delta s_i.
\label{redsum}
\end{equation}
We consider a nova to be detected if it has an apparent R-band magnitude less than a threshold magnitude, $m_{R} < m_{th}$.

Reddening values were compared to those recently estimated from SDSS maps by \citet{Schlafly:2011}. For $\Delta m_R < 5$ in the  DJ model, we obtain an rms residual value of 2.30. For $\Delta m_R < 10$, this increases to 4.87. 

\subsection{$\gamma$-ray Properties}
It was assumed that all novae emit in $\gamma$-rays. For a nova to be defined as a $\gamma$-ray source, we require $TS > 25$ (equivalent to $5\sigma$ when modelled using a simple power law model) over the emission period. In order to assign each nova with a $\gamma$-ray luminosity, the 1-day bin peak values for the existing {\sl Fermi} LAT detected $\gamma$-ray novae were taken, and a flat distribution assumed between them. The simulated novae were each assigned a $\gamma$-ray luminosity based on this distribution. This was done as a nova is more likely to be detected in $\gamma$-rays when at its peak. Although V5568 Sgr lacks a daily flux with $TS > 25$, it was still detected overall to $>5\sigma$ and its peak flux was included as the possibility remains that most of the $\gamma$-ray novae are more luminous than average. This assignment required the use of nova distances, which as previously discussed can be unreliable, as such the allowed  $\gamma$-ray luminosity range was defined by the dimmest nova, V5668 Sgr, and the brightest V1324 Sco. Table \ref{nova_table} shows that the percentage uncertainty on the distance to each of these novae is $\approx 20\%$. As $L_{\gamma} \propto d^{2}$, and these manifest themselves as $\approx 40\%$ uncertainties in  $L_{\gamma}$ when combined with the $F_{\gamma}$ uncertainties. As such, we extend our luminosity range to account for these uncertainties, therefore our nova population will contain novae with intrinsic luminosities consistent with the range of those observed. 

In addition, the source in question must be visible against the sky background, which is described by the {\sl Fermi} LAT background models $\rm{gll\_iem\_v06.fits}$ (Galactic diffuse) and  $\rm{iso\_P8R2\_SOURCE\_V6\_v06.txt}$  (isotropic diffuse)\footnote{{\sl Fermi} background models can be downloaded from \url{http://Fermi.gsfc.nasa.gov/ssc/data/access/lat/BackgroundModels.html}}. To quantify this, the overall background flux from the Galactic diffuse, $F_{GalDiff}$, was taken for the pixel containing each detected $\gamma$-ray nova, and the ratios of peak daily flux (with $TS > 25$) to background flux, $F_{\gamma}/F_{GalDiff}$ were calculated. The effects of the isotropic diffuse were deemed insignificant due to the proximity of the $\gamma$-ray novae to the Galactic plane, and so was neglected. They are listed in Table \ref{nova_table}. An additional criteria for $\gamma$-ray detection was therefore that the ratio $F_{\gamma}/F_{GalDiff}$ was greater than the mean $F_{\gamma}/F_{GalDiff}$, namely $\left[F_{\gamma}/F_{GalDiff}\right]_{mean} =0.214$ for each simulated nova event. We therefore expect to see $\approx 6$ $\gamma$-ray detected novae for every 69 that are R-band visible, equivalent to $\approx 8.7\%$, if their apparent rarity is caused by proximity effects.

\section{Results}

Results are based on 100 simulations each of 8-year novae populations. Error bars are taken as the standard deviations of the 100 results, and so are quoted to $1\sigma$.

\begin{figure}
\centering
\begin{subfigure}[t]{\linewidth}
\includegraphics[width = 9.0cm, keepaspectratio=true]{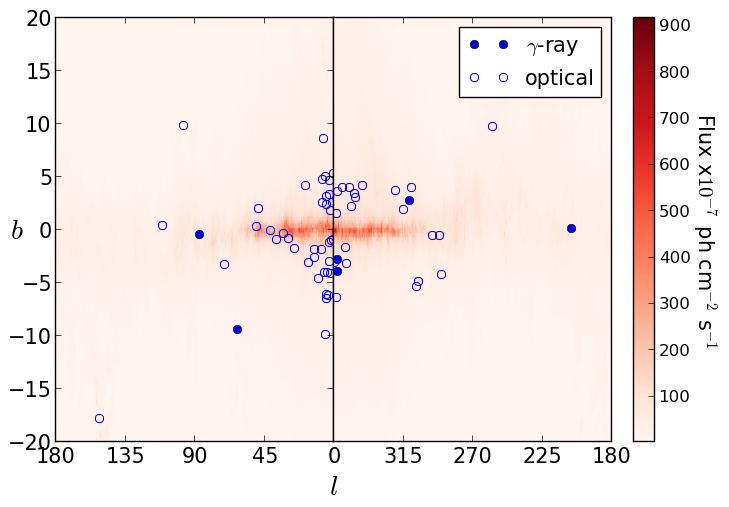}
\caption{Galactic Novae}
\end{subfigure}

\begin{subfigure}[t]{\linewidth}
\includegraphics[width = 9.0cm, keepaspectratio=true]{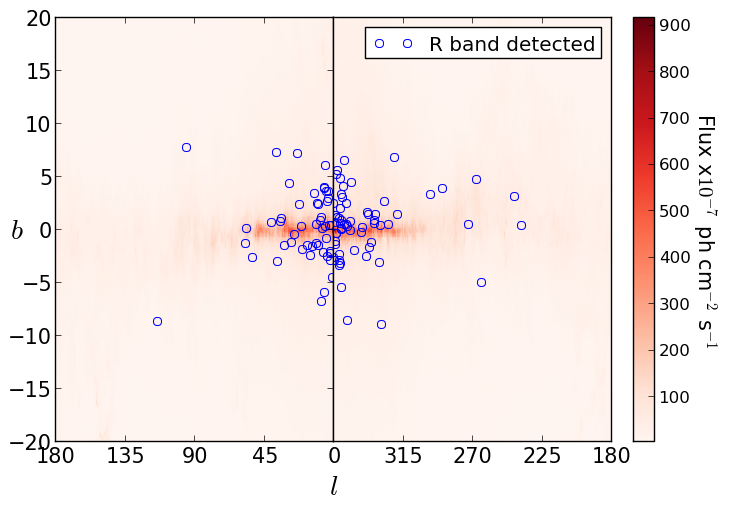}
\caption{Simulated Novae}
\end{subfigure}

\begin{subfigure}[t]{\linewidth}
\includegraphics[width = 9.0cm, keepaspectratio=true]{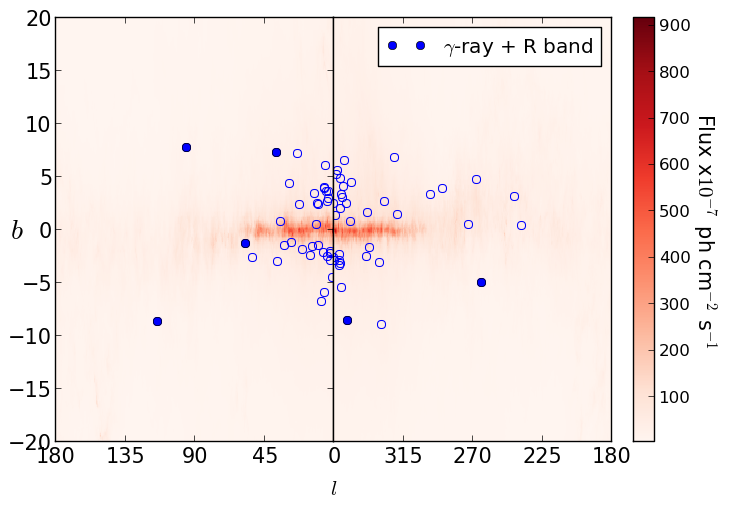}
\caption{Visible Novae}
\end{subfigure}
\caption{(a) Distribution of Galactic novae on the sky.  (b) Example simulated population. (c) Visible novae accounting for interstellar extinction in our simulated population. The colour scale represents the intensity of the $\gamma$-ray sky background.}
\label{novaeonsky}
\end{figure}

\begin{figure}
\includegraphics[width = 8.5cm, keepaspectratio=true]{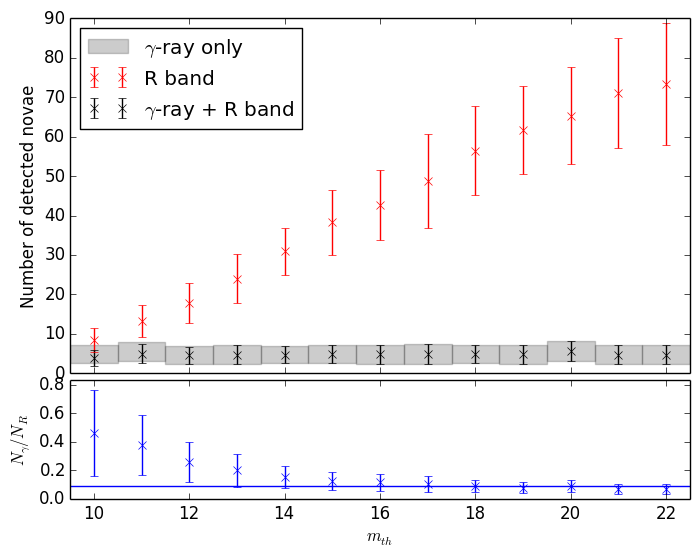}
\caption{Figure showing how the number of R-band and $\gamma$-ray detected novae vary as a function of $m_{th}$. It can be seen that for sufficiently dim $m_{th}$, the predicted ratio is consistent with the observed ratio which is shown by the horizontal line in the lower panel. The error bars are taken as the standard deviation based on 100 runs of the code.}
\label{ratio}
\end{figure}

We find that our model is best able to reproduce the correct number of observed novae when the global nova rate is $\dot{N}_{novae} \approx 20~ \rm{year}^{-1}$. Fig. \ref{novaeonsky} demonstrates the success of our model to reproduce the observed distribution of novae, and show that interstellar extinction effects are greatest when observing through the Galactic plane towards the Galactic centre, thus coinciding with the region of the highest $\gamma$-ray background. This implies that the population of novae in the Milky Way is bulge dominated, much like in M31.  Fig. \ref{ratio} shows that our simulated population produces a fraction of novae consistent with  observations for any limiting R-band magnitude with $m_{th}<13$. As novae in our M31 sample were as dim as $m_{R} = 20.6$ , this is strong evidence to say such a rudimentary model can reproduce the observed $\gamma$-ray nova fraction, validating our assumptions. The fact that the number of $\gamma$-ray novae is consistent with being constant across the range of $m_{th}$ values implies that the $\gamma$-ray sky background flux is the dominant factor prohibiting the discovery of further $\gamma$-ray novae. It can also be seen that at low $m_{th}$ a $\gamma$-ray nova is more likely to be observed lacking an optical counterpart. Typically there was one per simulation, so the unidentified $\gamma$-ray sources in the {\sl Fermi} 3FGL catalogue are unlikely to contain many novae. 


\begin{figure}
\includegraphics[width = 8.5cm, keepaspectratio=true]{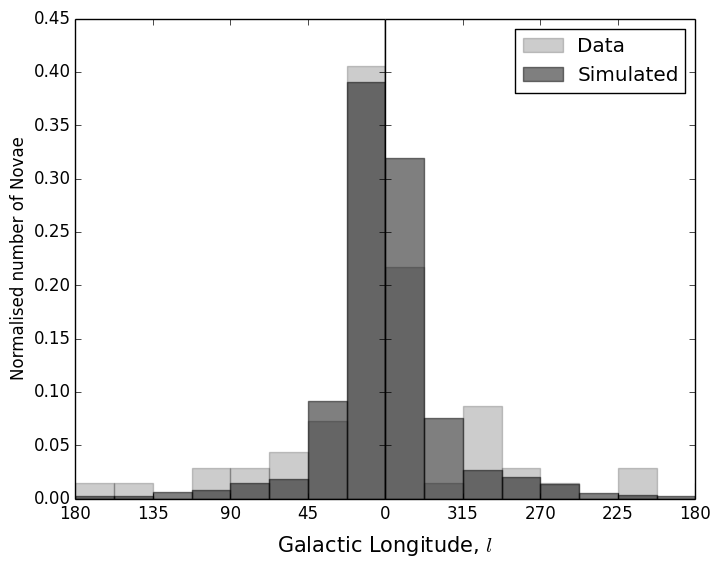}
\caption{Comparison of the $l$ values for Galactic novae to simulated novae, where the simulated novae have been assigned $l$ values based on the spatial binning of M31 novae in Fig. \protect\ref{EllipticalBins} scaled to a Milky Way radius of 20 kpc. Although to a lesser extent than in the data, our model reproduces the asymmetry in the Galactic longitude distribution, which we attribute to the Solar System being closer to the bulge at $45>l>0$ relative to $360>l>315$.}
\label{ldistr}
\end{figure}

The axial symmetry of our assumed nova distribution is shown in Fig. \ref{ldistr}. It can be seen that, like the observed novae, the simulated novae have a larger population at $45>l>0$ relative to $360>l>315$, albeit to a lesser extent. We attribute this to the Solar System being closer to the Milky Way bulge at the smaller $l$ values.

Fig. \ref{mRvdist} illustrates the range of $m_R$ values for the simulated novae as measured from Earth. It demonstrates that a large fraction of the total novae are far too dim to be observed realistically and that only CNe with $m_{R} \leq 12$ and within $\approx 8~\rm{kpc}$ are likely to be discovered in $\gamma$-rays.

The assumed power law distribution of novae $\gamma$-ray luminosities as a function of distance is displayed in Fig. \ref{Lvsdist}. It can be seen that the effect of our assumed flat distribution is to broaden the effective index of the overall spectrum, and that all novae within $\approx 7 ~\rm{kpc}$ with $F_{\gamma} > 5 \times 10^{-7}~\rm{photons~s}^{-1}~\rm{cm}^{-2} $ are discovered both in $\gamma$-rays and the R-band, which is consistent with observations. The figure indicates that we should be able to optically confirm the majority of novae within $\approx 7 ~ \rm{kpc}$ from us.

\section{Discussion}

As this study has been based on observed novae in M31, any intrinsic differences between the M31 and MW novae must be discussed. It is clear from Fig \ref{MagBins} that M31 has a bulge dominated nova population, which is something we cannot directly confirm for the MW due to our location within the disc and the aforementioned difficulties in measuring novae distances, which impose restrictions on our ability to deduce this. It has been hypothesised that barred spiral galaxies can drive star formation in galactic centres, and \citet{Beaton2006} use near infrared data for M31 to conclude that it is a barred spiral much like our Milky Way. The bar can transfer gas and drive star formation in the bulge, thus leading to a higher stellar population than in the disc, and therefore more binary systems, some of which will be CVs capable of producing nova outbursts. This can explain the observed bulge-dominated population, and the similarity of M31 to the MW suggests the MW nova population need also be bulge dominated and justifies our use of M31 novae. It is unlikely that the M31 nova population is an observational artefact caused by reddening effects exclusive to the disc, and we reproduce the distribution of MW novae on the sky implying a bulge dominated population, contradicting the findings of \citet{1997ApJ...487L..45H} who find M31 has a disc dominated nova population.

We find that adjusting the bulge to disc nova fraction has very little effect on overall nova rates in either waveband, but a larger affect on the sky distribution, whereby the simulated coordinates diverge from those observed. This is a consequence of the binning criteria as disc bins closer to the Galactic centre contain more novae. Therefore the effect of reducing the proportion of bulge novae is effectively to shift them to the inner bin of the disc, where those on the near side to the solar system are mostly observed and those further away are not. It is clear that the bin immediately surrounding the bulge contains the largest errors which are not taken into account, both from bulge related projection effects in obtaining the M31 distribution and defining a definite bulge boundary. Strong reddening and $\gamma$-ray background levels in these regions mitigates these effects to the extent that making the MW bulge semi major axis $2~\rm{kpc}$ has very little impact on the number of novae that can be detected. 

Fig. \ref{ldistr} implies M31 and MW novae are distributed in a similar manner, and 2 sample KS tests on the output simulated distribution give a $\approx 50\%$ chance that the distributions can arise from samples of the same global population. The discrepancies arise at points far from the Galactic centre, in regions of low interstellar extinction, implying novae need be slightly more spread out in the Milky Way relative to Andromeda. This could be because our simulated Milky Way is smaller than M31 ($R_{M31} = 27~\rm{kpc}$ \citep{1958M31size}), but a more likely explanation is that our nova sample is not large enough. For a complete sample, we would not expect to see empty $l$ bins, though depleted bins could be indicative of areas with higher interstellar extinction. Due to difficulties in measuring reddening effects, we conclude that any lack of longitudinal symmetry exhibited by observed Galactic novae instead highlights the difficulties in modelling Galactic reddening, and that reddening effects need not be symmetric about the MW centre. Furthermore, novae can occur in any region on the sky, hence regions with preferential sky coverage are likely to contain more novae, imposing a bias on our data set.

\begin{figure}
\includegraphics[width = 8.5cm, keepaspectratio=true]{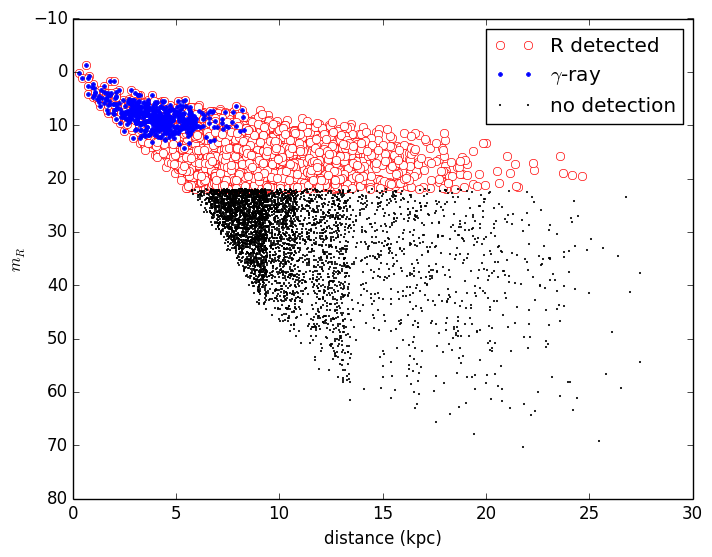}
\caption{Figure demonstrating the detectability of novae with varying $m_{R}$ as a function of distance. In this case, $m_{th} = 22$.}
\label{mRvdist}
\end{figure}

\begin{figure}
\includegraphics[width = 8.5cm, keepaspectratio=true]{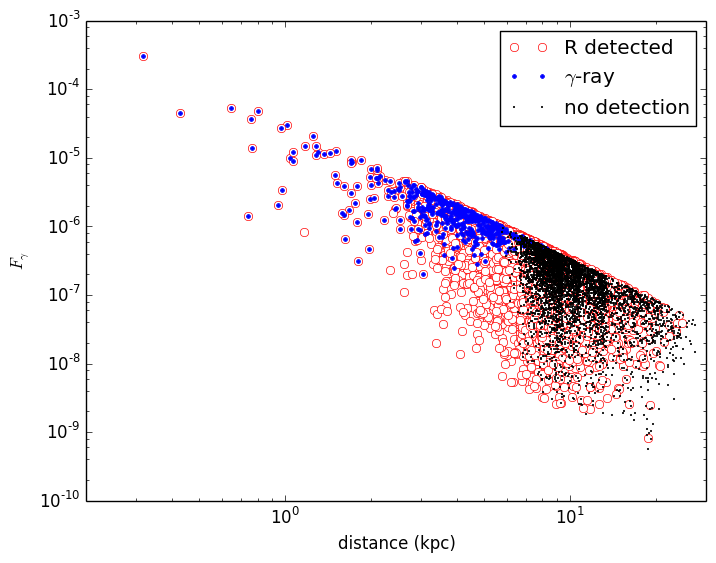}
\caption{Figure demonstrating the detectability of novae with varying $m_{R}$ as a function of distance. In this case, $m_{th} = 22$. It can be seen that the assumed power law relationship is recovered, and that the majority of novae detected within 5 kpc are detected both in $\gamma$-rays and in the R-band.}
\label{Lvsdist}
\end{figure}

Figures \ref{mRvdist} and \ref{Lvsdist} show the range of values for $m_{R}$ and $F_{\gamma}$ when both are a function of distance. These essentially explain how $\gamma$-ray and R-band light propagate through our simulated Milky Way. The large spread for $m_{R}$ is a direct manifestation of the range of interstellar extinction values experienced by novae with significantly different line of sight paths to the Solar System. In contrast, the spread in $F_{\gamma}$ only reflects that of the defining population, hence the index is reasonably approximated by a power law. This again highlights the importance of accurately determining interstellar extinction. Fig. \ref{Lvsdist} exhibits a sudden cutoff in detectability (blue to black transition) occurring between $d \approx 6-7~\rm{kpc}$. This is because in our simulations the Solar System is located $8~\rm{kpc}$ from the Galactic centre and $\approx 6~\rm{kpc}$ from the nearest point of the elliptical bulge, and novae in this region are both more likely to be dominated by the $\gamma$-ray sky background and experience stronger extinction effects, rendering them undetectable.

Despite the importance of interstellar extinction with respect to nova discovery, the $\gamma$-ray sky background completely dominates when attempting to discover $\gamma$-ray novae. Fig \ref{novaeonsky} demonstrates that novae observed in $\gamma$-rays are typically not close to the Galactic plane, something which our simulated population in Fig. \ref{novaeonsky} reproduces. This is a direct consequence of the $\gamma$-ray background being significantly smaller further from the Galactic plane. One particular consequence of this is that any optically unassociated objects in the {\sl Fermi} LAT catalogue \citep{2015LATcat} are unlikely to be classical novae. Novae observed at high $|{l}|$ are more likely to be nearby simply for geometrical reasons, less likely to suffer optically from interstellar extinction and more likely to be discovered in $\gamma$-rays due to the lower background. This combination of facts essentially explains the relative ease with which our rudimentary model can reproduce the fraction of $\gamma$-ray novae. 

With regards to observing a CNe in both $\gamma$-rays and the R-band, Figures \ref{mRvdist} and \ref{Lvsdist} are of particular interest. The blue region (dark grey in printed version) in Fig. \ref{mRvdist} indicates that we can only realistically expect to detect novae in $\gamma$-rays for $m_R < 12$ and $d \leq 8~\rm{kpc}$, with the majority of these within $6~\rm{kpc}$. These figures be used to explain that the non detection of Nova KT Eri (distance $ 6.3 \pm 0.1~\rm{kpc}$ \citep{2013KTEri}, $m_{V} = 8.1$ \citep{2009CBET.2050....1Y}) is as a result of the nova being less luminous in $\gamma$-rays that those discovered. \citet{2013KTEri} also discusses the possibility of KT Eri being a recurrent nova, and hence may not belong to the same class of objects. Again, these numbers are the manifestation of parameters in our model. Looking in the $l=0$ direction, novae can only be detected optically and in $\gamma$-rays away from the Galactic plane. Whilst optical novae trace Galactic reddening, $\gamma$-ray fluxes follow an inverse square law and so only the more luminous novae can be observed further away than $6~\rm{kpc}$. Even then, they need to be located in a region of low enough $\gamma$-ray background, which is unlikely given the bulge-dominated spatial distribution. Neglecting off plane effects, this represents  $\approx (6~\rm{kpc}/20~\rm{kpc})^2 = 9\%$ fraction of our Galaxy, which is close to the observed $8.7\%$ of CNe detected in $\gamma$-rays. This simple argument supports the fact that $\gamma$-ray novae are rare only because they need to be close by to be detected.

The number of identifiable  $\gamma$-ray detectable novae is independent of $m_{R,th}$, and Fig. \ref{ratio} illustrates that the ratio $N_{\gamma}/N_R$ decreases with increasing $m_{R, th}$. Whilst $N_{\gamma}/N_R$ can be tweaked by the number of R-band novae visible, $N_{\gamma}$ is always consist with observations, and depends only on the global nova rate, which is optimised at $\dot{N}_{novae} \approx 20~ \rm{year}^{-1}$. Although lower than the inferred rate of \citet{Shafter:1997}, we deem our conclusions still valid as the goal was to reproduce the observed nova population on the sky, and from that draw conclusions about the number of $\gamma$-ray novae. The same argument applies to our redding parameters, where we use $z_d = 0. 2~\rm{kpc}$ instead of the original $z_d = 0.1~\rm{kpc}$ used by \citet{DJ:1994}. This was necessary to avoid a large population of novae in the range $ 2> |b| > 0$ which is not observed.

Referring to Table \ref{nova_table}, it is clear that V1324 Sco, V1369 Cen and V5668 Sgr were detected with the LAT with ratios $F_{\gamma}/F_{GalDiff} < \left[F_{\gamma}/F_{GalDiff}\right]_{mean} =0.214$. Therefore it is possible that novae in our simulations not considered detectable at $\gamma$-ray energies would have indeed been detected by {\sl Fermi}, therefore increasing our ratio $N_{\gamma}/N_{R}$. On average, this effect would cancel with those with $F_{\gamma}/F_{GalDiff} > \left[F_{\gamma}/F_{GalDiff}\right]_{mean} =0.214$ located in regions of the sky with high background fluxes. Such an event can be attributed to our simulated $\gamma$-ray luminosities being based off a sample of only six novae. V5668 Sgr is of particular interest because it implies that the nova was intrinsically fainter in $\gamma$-rays than the others. Transient phenomena are always subject to a bias favouring those events which are more luminous due to their ease of discovery and study. Thus, our simulated $\gamma$-ray population may be more luminous on average than the global population, assuming all novae do emit $\gamma$-rays. If this were the case, we would expect to see fewer $\gamma$-ray novae reducing our $N_{\gamma}/N_{R}$. Clearly any future studies on $\gamma$-ray novae would benefit from a larger source sample size, which would give insight into the number of novae per unit energy and could replace the assumed flat distribution.

\section{Conclusions}

Assuming Milky Way novae are similar in R-band magnitude and spatial distribution to M31, a population of novae was simulated over the first 8 years of {\sl Fermi} LAT observation time, during which 6 out of 69 have been detected in $\gamma$-rays. This was done by dividing M31 into 2 bulge and 4 disc spatial bins, and binning R-band magnitudes of novae for both bulge and disc. Simulated novae were assigned R-band peaks based on their spatial location (disc or bulge) in the Milky Way, with Milky Way spatial bins for the disc and bulge separately normalised such that they contain the same fractional areas as their M31 counterparts. M31 nova rates were computed per unit area on the sky, and scaled to the Milky Way, allowing a simulated Galactic nova population to be produced. We assumed a Galactic disc of radius $R_{MW} = 20~\rm{kpc}$ and a bulge with semi-major axis $a=3.0~\rm{kpc}$ with 2D axis ratios 2:1.

The spatial locations of simulated novae were converted to galactic coordinates. The longitude was done geometrically, whereas the latitude form disc novae assumed exponential decay profiles of scale heights $z_d = 350~\rm{pc}$, whilst bulge novae were found to best follow a Gaussian profile, $\rho_{1} = \rho_0 \exp(-0.5 r^2)$,  with $r = [ \left({x}/{x_0} \right)^2 + \left({y}/{y_0} \right)^2 + \left({z}/{z_0} \right)^2 ]^{0.5}$ and best fit parameters  $\rho_{0}=1 \times 10^6$,  $x_0=4.17$,  $y_0=0.674$ and  $z_0=0.344$. Optically, the double exponential disc extinction model of \citet{DJ:1994} was assumed, allowing the total amount of reddening in the R-band along the line of sight to be determined. This yielded a $m_R$ value for each nova, which if was smaller than the free parameter $m_{R, th}$, led the nova to be classed as discoverable in the R-band.

Simulated novae were assigned $\gamma$-ray peaks based on a flat distribution of 24 hour bin maximum $TS$ values for the existing novae light curves and assuming an inverse square law relationship between $\gamma$-ray peak and distance the $\gamma$-ray flux was calculated at the Earth. This was then compared to the average $\gamma$-ray background flux at the location on the {\sl Fermi} LAT all sky map consistent with the location of each nova. If the nova flux was greater than the threshold of $\left[F_{\gamma}/F_{GalDiff}\right]_{mean} =0.214$, it was recorded as a detection in $\gamma$-rays. 

We find that for all values of $m_{th}$, the number of novae observable in $\gamma$-rays, $N_{\gamma}$, is consistent with the number both observable in $\gamma$-rays and the R-band, with only small exceptions present for small $m_{th}$. We attribute this to the $\gamma$-ray background being the most significant hindrance to the discovery of $\gamma$-ray novae. Our simulations tell us that any given nova is unlikely to be discovered in $\gamma$-rays if $m_{R} \geq 12$ and $d > 8~\rm{kpc}$, and that the ratio $N_{\gamma}$/$N_R$ is consistent with the observed ratio for all $m_{R, th} < 13$. This demonstrates that observed nova rates can easily be reproduced with sensible parameters from a simple model, implying that $\gamma$-ray novae are indeed nearby rather than intrinsically rare phenomena.

\section*{Acknowledgements}

This work was supported by the Oxford Centre for Astrophysical Surveys which is funded through generous support from the Hintze Family Charitable Foundation. GC acknowledges support from STFC grants ST/N000919/1 and ST/M00757X/1 and from Exeter College, Oxford. We would also like to thank Retha Pretorius and Matt Ridley for useful discussions about novae and galactic dynamics, and the Department of Physics at Durham University.




\bibliographystyle{mnras}

\bibliography{GamRayNovaeRarity_bib}







\bsp	
\label{lastpage}
\end{document}